\documentclass{aastex}
\usepackage{spr-astr-addons}
\usepackage{url}

\RequirePackage{color}

\begin{document}

\title{WD + MS systems as the progenitor of SNe Ia}
\shorttitle{WD + MS for SNe Ia} \shortauthors{Meng \& Yang}

\author{Xiangcun Meng\altaffilmark{1}}
\and \author{Wuming Yang\altaffilmark{1}} 

\altaffiltext{1}{Department of Physics and Chemistry, Henan
Polytechnic University, Jiaozuo, 454000, China\\E-mail:
xiangcunmeng@hotmail.com}

\begin{abstract}
The single-degenerate (SD) channel for the progenitors of type Ia
supernovae (SNe Ia) is one of the most popular channels, in which
a carbon-oxygen white dwarf (CO WD) accretes hydrogen-rich
material from its companion, increases its mass to the
Chandrasekhar mass limit, and then explodes as a SN Ia. We show
the initial and final parameter space for SNe Ia in a ($\log
P^{\rm i}, M_{\rm 2}^{\rm i}$) plane and find that the positions
of some famous recurrent novae, as well as a supersoft X-ray
source (SSS), RX J0513.9-6951, are well explained by our model.
The model can also explain the space velocity and mass of Tycho G,
which is now suggested to be the companion star of Tycho's
supernova. Our study indicates that the SSS, V Sge, might be the
potential progenitor of supernovae like SN 2002ic if the delayed
dynamical-instability model due to Han \& Podsiadlowski (2006) is
appropriate. Following the work of Meng, Chen \& Han (2009), we
found that the SD model (WD + MS) with an optically thick wind can
explain the birth rate of supernovae like SN 2006X and reproduce
the distribution of the color excess of SNe Ia. The model also
predicts that at least 75\% of all SNe Ia may show a polarization
signal in their spectra.

\end{abstract}


\section{Introduction}
Type Ia supernovae (SNe Ia) play an important role as cosmological
distance indicators to measure cosmological parameters (e.g.
$\Omega$ and $\Lambda$; Riess et al. 1998; Perlmutter et al.
1999), which have led to the discovery of the accelerating
expansion of the universe. However, the exact nature of SNe Ia
progenitors has not been clear (see the reviews by Hillebrandt \&
Niemeyer 2000; Leibundgut 2000). It is widely accepted that a SN
Ia originates from the thermonuclear runaway of a carbon-oxygen
white dwarf (CO WD) in a binary system. The CO WD accretes
material from its companion, increases mass to its maximum stable
mass, and then explodes as a thermonuclear runaway (Branch 2004).
Among all the progenitor models, the single-degenerate (SD)
Chandrasekhar model is widely studied (Yungelson et al. 1995; Li
\& van den Heuvel 1997; Hachisu et al. 1999a, 1999b; Nomoto et al.
1999; Langer et al. 2000; Han \& Podsiadlowski 2004; Chen \& Li
2007; Han 2008; Meng, Chen \& Han 2009; L\"{u} et al. 2009) and
upheld by many observation, e.g. the discovery of the potential
companion of Tycho's supernova (Ruiz-Lapuente et al. 2004) and the
variable circumstellar absorption lines (Patat et al. 2007). In
the SD model, the maximum stable mass of the CO WD is $\sim 1.378
M_{\odot}$ (close to the Chandrasekhar mass, Nomoto, Thielemann \&
Yokoi 1984), and the companion is probably a main sequence star or
a slightly evolved star (WD+MS), or a red-giant star (WD+RG)
(Whelan \& Iben 1973; Nomoto, Thielemann \& Yokoi 1984). In the
paper, we focus on the WD+MS channel, which is a very important
channel to contribute to SNe Ia.

The paper is organized as follows: we describe our methods in
Section 2, show our results in Section 3, present some discussions
in Section 4, and then finally in Section 5, we give our
conclusions.

\section{Methods}
Our method for treating the binary evolution of WD + MS systems is
the same as that in Meng, Chen \& Han (2009), and here, we only
give a simply description. We use the stellar evolution code of
Eggleton (1971, 1972, 1973) to calculate the binary evolutions of
WD+MS systems. Instead of solving the stellar structure equations
of a WD, we adopt the prescription of Hachisu et al. (1999a) on
WDs accreting hydrogen-rich material from their companions. We
assume that if the mass-transfer rate, $|\dot{M}_{\rm 2}|$,
between WD and its companion is larger than a critical value,
$\dot{M}_{\rm cr}$, the accreted hydrogen-rich material steadily
burns on the surface of WD at the rate of $\dot{M}_{\rm cr}$,
while the unprocessed matter is lost from the system as an
optically thick wind at a rate of $\dot{M}_{\rm
wind}=|\dot{M}_{\rm 2}|-\dot{M}_{\rm cr}$ (Hachisu et al. 1996).
The material lost as the optically thick wind may exist as
circumstellar material (CSM) and be a possible origin of color
excess of SNe Ia (Meng et al. 2009). When $|\dot{M}_{\rm 2}|$
locates in the range of $\dot{M}_{\rm cr}$ to
$\frac{1}{8}\dot{M}_{\rm cr}$, the hydrogen burning is stable or
weakly unstable and no material is lost from the system. If
$|\dot{M}_{\rm 2}|$ is smaller than $\frac{1}{8}\dot{M}_{\rm cr}$,
the hydrogen burning is heavily unstable and no material
accumulates on the surface of the WD. The hydrogen-rich material
accumulated is converted into helium. When the mass of the helium
reaches a certain value, a helium flash may occur and a part of
helium may be lost from the system. The mass accumulation
efficiency for helium-shell flashes is from Kato \& Hachisu
(2004). If the mass of a WD can reach 1.378 $M_{\odot}$, we assume
that the WD explodes as a SN Ia.

\section{Results}
\subsection{initial and final parameters space}
We carried out a detailed binary evolution calculation for more
than 25,000 WD+MS systems with various metallicities, and we
obtained a large, dense model grid. The final outcomes of all the
binary evolution calculations are summarized in the initial
orbital period-secondary mass ($\log P^{\rm i}, M_{\rm 2}^{\rm
i}$) plane (see Figs. 2 and 3 in Meng, Chen \& Han 2009).

\begin{figure}[t]
\begin{center}
\includegraphics[width=6.2cm,angle=270]{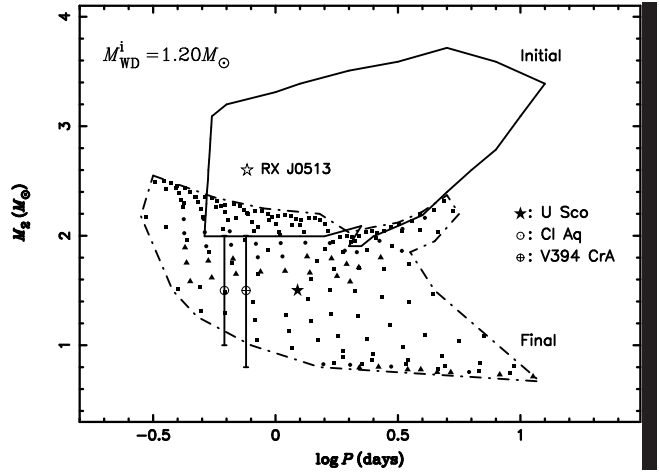}
\caption{Parameter regions producing SNe Ia in the ($\log P-M_{\rm
2}$) (orbital period-donor mass) plane for the WD + MS systems.
The initial WD mass is $1.20 M_{\odot}$. The WD + MS system
insides the region encircled by solid line (labeled ``initial'')
will increase its white dwarf mass up to 1.378 $M_{\odot}$, where
we assume a SN Ia explosion. The final state of the WD + MS system
in the plane is encircled by a dot-dashed line (labeled
``final''). Filled squares indicate SN Ia explosions during an
optically thick wind phase. Filled circles and filled triangles
denote SN Ia explosions after the wind phase while hydrogen-shell
burning is stable or mildly unstable, respectively. A supersoft
X-ray source, RX J0513.9-6951, (open star) is plotted. Three
recurrent novae are indicated by a filled star, a solar symbol and
an earth symbol, respectively.} \label{mms02120}
\end{center}
\end{figure}

Our results may provide help to judge whether a WD + MS system may
explode as a SN Ia. In Fig. \ref{mms02120}, we show the initial
contour for SNe Ia and the final state of binary evolution in the
($\log P-M_{\rm 2}$) plane at the moment of SNe Ia explosion for
the case of $M_{\rm WD}^{\rm i}=1.2M_{\odot}$ with $Z=0.02$. The
cases of other WD masses and other metallicities are similar. In
the figure, we also show the position of a supersoft X-ray source
(SSS, RX J0513.9-6951) and three famous recurrent novae, i.e. U
Sco, V394 CrA and CI Aql. Considering the WD masses of RX
J0513.9-6951 and U Sco and their positions in the ($\log P-M_{\rm
2}$) plane, they are very likely candidates for the progenitors of
SNe Ia. Due to the uncertainties of the companions of V394 CrA and
CI Aql, they are potential candidates for the progenitors of SNe
Ia (See Hachisu et al. (2008) and Meng, Yang \& Geng (2009) for
similar discussions about these objects).

\subsection{companion properties}
\begin{figure}[t]
\begin{center}
\includegraphics[width=6.2cm,angle=270]{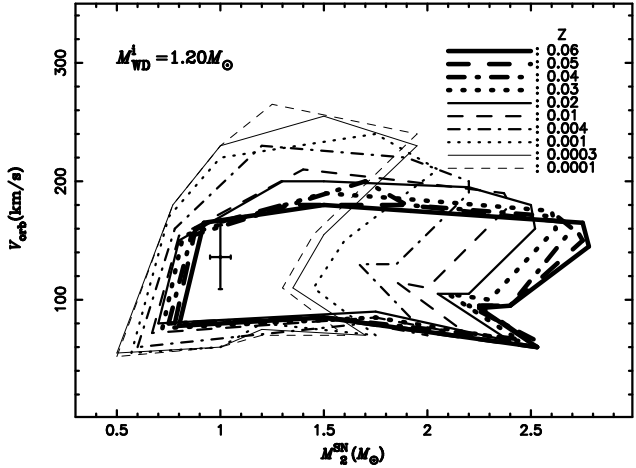}
\caption{The final states of companion stars at the moment of SNe
Ia explosion for different metallicities. The initial WD mass is
$1.20M_{\odot}$. Cross represents Tycho G, which is a potential
candidate for the companion of Tycho's supernova (Ruiz-Lapuente et
al. 2004; Branch 2004). The length of the cross represents
observational errors.} \label{mvz}
\end{center}
\end{figure}

We also show the parameter spaces for companions at the moment of
supernova explosion. In Fig. \ref{mvz}, we show the final states
of the companions with various metallicities at the moment of
supernova explosion in the ($M_{\rm 2}^{\rm SN}-V_{\rm orb}$)
(final companion mass - orbital velocity) plane. We noticed that
the suggested companion star of Tycho's supernova, Tycho G,
locates in the permitted region and all metallicities seem to
account for the position of the potential companion. Actually, the
detailed binary population synthesis results show that only the
case of $Z=0.02$ is appropriate for Tycho G (see also Meng, Yang
\& Geng 2009).

\subsection{SN 2002ic}
SN 2002ic is the first object showing hydrogen lines in its
spectrum, which is interpreted as the interaction between
supernova ejecta and CSM (Hamuy et al. 2003). Many models have
been suggested to explain this rare object. Among all the models,
the delayed dynamical-instability model suggested by Han \&
Podsiadlowski (2006) is more interesting and some predictions from
the model seem to be consistent with observations, especially in
the sense of the birth rate and delay time of the rare object
(Aldering et al. 2006; Prieto et al. 2007). In the scenario of Han
\& Podsiadlowski (2006), SN 2002ic may be from the WD + MS
channel, where the CO WD accretes material from its relatively
massive companion ($\sim 3.0M_{\odot}$), and increases its mass to
1.30 $M_{\odot}$ before experiencing a delayed dynamical
instability. Following the study of Han \& Podsiadlowski (2006),
Meng, Chen \& Han (2009) showed the parameter space for SN 2002ic
with various metallicities and predicated that supernovae like SN
2002ic may not be found in extremely low-metallicity environments.
However, more evidence is necessary to confirm the scenario
suggested by Han \& Podsiadlowski (2006), especially to find a
progenitor system.

V Sge is a well observed quasi-periodic transient SSS in our
Galaxy. We found that it just locates at the boundary of
parameters for SN 2002ic in ($\log P^{\rm i}, M_{\rm 2}^{\rm i}$)
plane. The mass-loss rate for V Sge can be as large as
$\sim10^{\rm -5}M_{\odot}{\rm yr^{\rm -1}}$, indicated by radio
observation (Lockley et al. 1997, 1999), which is consistent with
the prediction from the delayed dynamical-instability model. So,
it is possible that V Sge is the first candidate of the progenitor
of supernovae like SN 2002ic if the delayed dynamical-instability
model in Han \& Podsiadlowski (2006) is appropriate. Based on the
prediction from Han \& Podsiadlowski (2006) that 1 in 100 SNe Ia
belongs to the subgroup of 2002ic-like supernovae and considering
that the life time of V Sge is $\sim10^{\rm 5}$ yr, there should
be several V Sge-type stars belonging to the WD + MS system in our
Galaxy if we take the Galactic birth rate of SNe Ia as
3-4$\times10^{\rm -3}{\rm yr^{\rm -1}}$ (van den Bergh \& Tammann
1991; Cappellaro \& Turatto 1997). Observationally, Steiner \&
Diaz (1998) listed four V Sge-type stars in the Galaxy and
discussed their similar spectroscopic and photometric properties.
So, the number of  V Sge-type stars is also consistent with the
prediction from the delayed dynamical-instability model (see Meng,
Yang \& Geng 2009 for details).

However, in our model grid, the case of $M_{\rm WD }^{\rm
i}=1.2M_{\odot}$ is the only one to account for the position of V
Sge in the ($\log P-M_{\rm 2}$) plane, but it currently seems that
$1.2M_{\odot}$ CO WDs are probably only in special circumstance to
form such as an extremely high metallicity (Meng, Chen \& Han
2008). In addition,  the CO WD with a mass high than
$1.1M_{\odot}$ seems not to explode as a SN Ia (Umeda et al.
1999). So it is premature to obtain a definitive conclusion that
there is a relation between the V Sge-type stars and the delayed
dynamical instability model.

\subsection{the birth rate}
Incorporating our results into Hurley¡¯s rapid stellar evolution
code (Hurley et al. 2000, 2002), we calculate the birth rate of
SNe Ia for the WD+MS model. We use Monte Carlo simulation to
generate a primordial binary sample (see Meng, Chen \& Han 2009
for the basic parameters for Monte Carlo simulation). We found
that for the case of single star burst, most supernovae occur
between 0.1 Gyr and 2 Gyr after star formation, and a high
metallicity leads to a systematically earlier explosion time. The
peak value of the birth rate increases with metallicity $Z$.
However, the WD+MS model only can account for about 30 percent of
the Galactic birth rate (see also Meng, Chen \& Han 2009).
\subsection{SN 2006X}
SN 2006X is the first case to show the signal of circumstellar
material, which shows its single-degenerate nature (Patat et al.
2007). Patat et al. (2007) suggested that the progenitor of SN
2006X is a WD + RG system based on the expansion velocity of the
circumstellar material.  Blondin et al. (2009) found that 6 in 100
SNe Ia should belong to the rare group. Hachisu et al. (2008)
suggested that SN 2006X may be from a WD + MS system, and if a WD
explodes at the optically thick wind phase, the SN may show a
signal like SN 2006X. Following the study of Meng, Chen \& Han
(2009), we calculated the evolution of the birth rate of
supernovae like SN 2006X by assuming that if a SN Ia explodes at
the optically thick wind phase, it is a 2006X-like supernova. We
found that the progenitor age of SN 2006X is between 0.3Gyr and
1Gyr, which means that there is star formation during the recent 1
Gyr in the host galaxy of supernovae like SN 2006X. The host
galaxies of SN 2006X and its twins, SN 1999cl, are both spiral
galaxies, and show a significant signal of star formation at
present (Kanpen et al. 1993, 1996; Wong \& Blitz 2002). We also
noticed that supernovae like SN 2006X are a relatively rare
subclass of SNe Ia: 1 to 14 in 100 SNe Ia can be of this type,
which depends on the common envelope ejection efficiency,
$\alpha_{\rm CE}$. The birth rate found by Blondin et al. (2009)
if located in the range. So, our results uphold the suggestion of
Hachisu et al. (2008) that the progenitor system of a supernova
like SN 2006X may be a WD + MS system (see Meng, Yang \& Geng
2009b).

\subsection{color excess}
In our binary evolution calculation, we assume that a part of the
hydrogen-rich material may be lost from the binary system as an
optically thick wind. The lost material may exist as circumstellar
material (CSM). If this scenario is appropriate, the CSM should be
the origin of the color excess of SNe Ia. Reindl et al. (2005)
showed the color excesses to be more than one hundred SNe Ia at
maximum light, which provides an opportunity to check whether the
SD model may reproduce the distribution of the color excess of SNe
Ia. Following the study of Meng, Chen \& Han (2009) and via a
simple analytic method, we may reproduce the distribution of color
excesses of SNe Ia by our binary population synthesis (BPS)
approach if the velocity of the optically thick wind is taken to
be of the order of 10 km s$^{\rm -1}$. However, if the wind
velocity is larger than 100 km s$^{\rm -1}$, the reproduction is
bad (see Meng et al. 2009 for details).

\subsection{Tycho G}
Following the study of Meng, Chen \& Han (2009) and via a BPS
approach, we calculated the distributions of the parameters of the
binary systems at the moment of supernova explosion and the
properties of companions after supernova explosion, e.g. mass,
radius, space velocity and surface gravity. The former may provide
physics input when one simulates the interaction between supernova
ejecta and its companion, and the latter may help to search for
the companions in supernova remnants. It is surprising that the
properties of the potential companion of Tycho's supernova, Tycho
G, are well consistent with our BPS results, and only the case of
$Z=0.02$ may account for the properties of Tycho G (please, see
our following paper: Meng \& Yang 2009). However, a conclusive
result about Tycho G is premature, since there have been
considerable debates about whether the star is related to Tycho's
supernova (Kerzendorf et al. 2009; Hern\'{a}ndez et al. 2009).

\subsection{polarization}
After the explosion, the explosion ejecta impacts into the
envelope of the companion and strips off some hydrogen-rich
material from the surface of the companion. Because of the
existence of the companion, a hole forms in the explosion ejecta
and the hole will never disappear since the movement of the ejecta
is supersonic. Then the explosion remnant is aspheric, and the
aspheric structure may reveal itself by a polarized spectrum
(Marietta et al. 2000; Meng, Chen \& Han 2007). Based on the
numerical simulation of Marietta et al. (2000) and Kasen et al.
(2004), we found that at least 75\% of all SNe Ia can be detected
by observation of the polarization. At present, all supernovae
which are observed by spectropolarimetry have various degrees of
polarization signal (Leonard et al. 2005).

\section{Discussion}
From our study, the WD + MS channel can only account for about
$1/3$ of the SNe Ia observed. Therefore, there may be other
channels or mechanisms contributing to SNe Ia. A wide symbiotic
system, WD + RG, is a possible progenitor of SNe Ia. Now, we are
constructing a complete model including WD + MS channel and WD +
RG channel, hoping to resolve the problem. In our new model, the
range of the companion mass for SNe Ia is from 0.6 $M_{\odot}$ to
5 $M_{\odot}$. The mass of WD contributing to SNe Ia may be as low
as 0.565 $M_{\odot}$.

An alternative is the double-degenerate (DD) channel. Although, it
is theoretically less favored, observation showed that it may
contribute to a few SNe Ia, e.g. SN 2003fg and SN 2006gz (Howell
et al. 2006; Hicken et al. 2007). So, at present it is premature
to obtain a definitive conclusion for the DD model.

\section{Conclusion}
The WD+MS channel is a very important channel for SNe Ia and the
channel may contribute to SNe Ia by about 30\%. Based on a
detailed binary evolution calculation, we show the initial and
final parameter space for SNe Ia in the ($\log P^{\rm i}, M_{\rm
2}^{\rm i}$) plane. The results may provide help to survey
potential candidates of SNe Ia. Following the results of Meng,
Chen \& Han (2009) and via a binary population synthesis approach,
we found that the SD channel may reproduce the birth rate of SN
2006X and the distribution of color excess of SNe Ia. The
properties of the potential companion of Tycho's supernova, Tycho
G, may also be consistent with our BPS results.

\end{document}